\documentclass[twocolumn,preprintnumbers,amsmath,amssymb]{revtex4}
\usepackage{dcolumn}
\usepackage{bm}
\usepackage{graphicx}
\usepackage{amsmath}
\usepackage{amsfonts}
\usepackage{amssymb}
\usepackage{amsthm}
\usepackage{amstext}
\usepackage{amsbsy}
\usepackage{amsopn}
\usepackage{amscd}
\usepackage{amsxtra}
\usepackage[colorlinks,bookmarksopen=false]{hyperref}

\hypersetup{%
  pdfstartview={FitH},
  linkcolor=blue,
  citecolor=blue,
  filecolor=black,
  menucolor=black,
  urlcolor =black
}

\def\phi{\varphi}
\def\epsilon{\varepsilon}

\def\ic{\textrm i}

\newcommand{\Exp}[1]{{\rm e}^{#1}}
\newcommand{\Sin}[1]{\sin\left({#1}\right)}
\newcommand{\Cos}[1]{\cos\left({#1}\right)}
\newcommand{\Tan}[1]{\tan\left({#1}\right)}

\newcommand{\Arccos}[1]{\arccos\left({#1}\right)}
\newcommand{\Arctan}[1]{\arctan\left({#1}\right)}

\newcommand{\bec}{\begin{center}}
\newcommand{\enc}{\end{center}}
\newcommand{\be}{\begin{equation}}
\newcommand{\ee}{\end{equation}}
\newcommand{\bmi}{\begin{minipage}}
\newcommand{\emi}{\end{minipage}}

\newcommand{\bi}{\begin{itemize}}
\newcommand{\ei}{\end{itemize}}
\newcommand{\ba}{\begin{array}}
\newcommand{\ea}{\end{array}}

\newcommand{\Der}[2]{\frac{\text{d}#1}{\text{d}#2}}

\newlength\figureheight
\newlength\figurewidth
\begin{document}
\title{%
  Time-optimal purification of a qubit in contact with a structured
  environment
}
\author{J. Fischer$^{1,2}$}
\author{D. Basilewitsch$^2$}
\author{C. P. Koch$^2$}
\author{D. Sugny$^{1,3}$}
\email{dominique.sugny@u-bourgogne.fr}

\affiliation{$^1$ Laboratoire Interdisciplinaire Carnot de Bourgogne (ICB), UMR
6303 CNRS-Universit\'e de Bourgogne- Franche Comt\'e, 9 Av. A. Savary, BP 47
870, F-21078 DIJON Cedex, FRANCE}
\affiliation{$^2$ Theoretische Physik, Universit\"at Kassel, Heinrich-Plett-Str.
40, D-34132 Kassel, Germany}
\affiliation{$^3$ Institute for Advanced Study, Technische Universit\"at
M\"unchen, Lichtenbergstrasse 2 a, D-85748 Garching, Germany}

\date{\today}

\begin{abstract}
  We investigate the time-optimal control of the purification of a qubit
  interacting with a structured environment, consisting of a strongly coupled
  two-level defect in interaction with a thermal bath. On the basis of a
  geometric analysis, we show for weak and strong interaction strengths that the
  optimal control strategy corresponds to a qubit in resonance with the
  reservoir mode. We investigate when qubit coherence and
  correlation between the qubit and the environment speed-up the control
  process.
\end{abstract}

\maketitle


\section{Introduction}
Controlling quantum systems with high efficiency in minimum time is of paramount importance for quantum
technologies~\cite{roadmapQT, Brif:10, Glaser:15, Altafini2012, Dong2010}. Since in
any realistic process, the system is inevitably subject to an interaction
with its environment, it is therefore crucial to understand the fundamental
mechanisms allowing to manipulate open quantum systems. A key point is the role that non-Markovianity (NM)~\cite{breuer:2016, vega:2017}
can play as resource for control~\cite{koch:2016}. Several studies have
recently pointed out the beneficial role of NM, for instance in the decrease of
quantum speed limit or in the protection of entanglement
properties~\cite{deffner:2013, campo:2013, mangaud:2018, Reich2015, poggi:2017,
mukherjee:2015}.

Quantum optimal control theory (OCT) is nowadays a mature field with
applications extending from molecular physics, nuclear magnetic resonance and
quantum information processing~\cite{Brif:10, Glaser:15, Altafini2012}.
A variety of numerical optimization algorithms has been developed so far to
realize different tasks~\cite{Werschnik2007, Reich2012, Bryson1975a, Doria2011a, Kelly2014, Khaneja2005, Krotov1996},
but also to account for experimental imperfections and
constraints~\cite{Borneman2010, Daems2013, Egger2014a,Kobzar2012, LTTS09}. Originally applied to closed quantum systems,
optimal control techniques have become a standard tool for open systems, both in
the Markovian and non-Markovian regimes~(see \cite{Glaser:15, koch:2016} and references therein). While OCT is very efficient and
generally applicable, it is often not straightforward to deduce the actual control mechanisms. In contrast, geometric and analytic optimal control techniques yield typically
more intuitive control solutions for low dimensional systems~\cite{Agrachev2004,DAlessandro2008, Jurdjevic1997}. Recent studies have shown
the potential of such methods both for closed~\cite{Albertini2015, Assemat2010,Boscain05, BM06, DAlessandro2001, GGS13, Khaneja2001,sugny:2008} and open quantum systems~\cite{Khaneja2003a, BCS09, Bonnard2012}. In this direction, while the control of
a dissipative qubit in the Markovian regime is by now well understood~\cite{LAGS13,
LZBGS10, MCA13, TB99}, very few studies have focused on
the case of a structured bath with a possibly non-Markovian
dynamics~\cite{Basilewitsch2017c, mukherjee:2015}. This is due to the inherent
complexity of such systems which prevents a geometric analysis.

In order to tackle this control problem, we consider a minimal
model of a controlled qubit coupled to a structured environment~\cite{Basilewitsch2017c}. The bath is
composed of a well-defined mode, a two-level quantum system (TLS), interacting
with a thermal reservoir, that can be described by a Markovian master equation. We assume that the external control field can only
modify the effective energy splitting of the qubit. The decisive advantage of
this simple control scenario is that a complete geometric and analytical
description can be carried out. We generalize Ref.~\cite{Basilewitsch2017c} where optimal control fields are
designed numerically and a geometric description is derived  when the interaction between the reservoir mode and the
thermal bath is neglected. The generalization allows us to analyze different configurations of the model
system geometrically for the whole range of parameters.

\begin{figure}[tb]
  \includegraphics[scale=1.0]{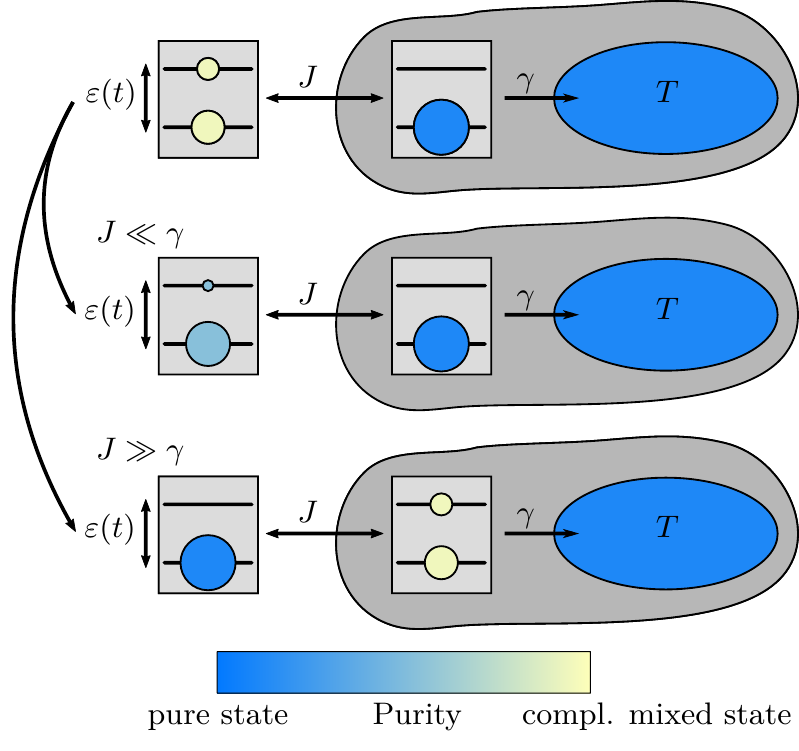}
  \caption{(Color online) Schematic representation of the purification setup: A qubit
    is coupled to an environmental TLS (with a coupling strength $J$) which
    decays with the rate $\gamma$ (defined in Eq.~\eqref{notations}) into a heat
    bath. In the weak coupling limit, which is identified to Markovian dynamics,
    the state of maximum purity cannot be reached in finite time (middle). If
    a strong coupling $J$ is considered then the qubit can be directed to the
    state of maximum purity (bottom).}
  \label{fig:setup}
\end{figure}

As an example control problem, we investigate the maximization of qubit purity in minimum time. Purification is a prerequisite in many applications. Qubit reset
has been shown through the coupling with a thermal bath~\cite{geerlings:2013,
valenzuela:2006, reed:2010, grajcar:2008}, but also by other
mechanisms~\cite{magnard:2018, jelezko:2004, johnson:2012, riste:2012,
riste:2012b}. A schematic description of the purification process used here is given in Fig.~\ref{fig:setup}. For the model system under study, we
analyze the interplay between NM, quantum speed limit and maximum available
purity. We show that the time-optimal reset protocol corresponds to a resonant
process for any coupling strength between the qubit and the TLS and decay rate
of the bath. We also discuss the role of initial coherences and correlations
between the qubit and the bath mode and we show that in some specific cases they allow to speed up the control and improve the final purity.

The remainder of this paper is organized as follows. The model system is
presented in Sec.~\ref{sec2}. A specific choice of coordinates allowing to
reduce the dimension of the control problem is described. Section~\ref{sec3} is
dedicated to the design of the time-optimal solution for the qubit purification
process. The role of initial coherences and correlations is discussed in
Sec.~\ref{sec4}. We conclude in Sec.~\ref{sec5}. Some technical formulas and
mathematical details are reported in Appendices~\ref{app:rwa}
and~\ref{app:coord}.

\section{Model}\label{sec2}
We consider a system consisting of a qubit whose effective energy
splitting $\omega_\text{q}$ can be modified by an external control field
$\varepsilon(t)$~\cite{Basilewitsch2017c}. The corresponding Hamiltonian reads
\begin{align} \label{eq:ham_q}
  \mathbf{H}_\text{q}(t) = -\frac{\omega_\text{q}}{2}
  \boldsymbol{\sigma}_\text{q}^z - \frac{\varepsilon(t)}{2}
  \boldsymbol{\sigma}_\text{q}^z,
\end{align}
where $\boldsymbol{\sigma}^x,\ \boldsymbol{\sigma}^y,\ \boldsymbol{\sigma}^z$
are the usual Pauli operators. The qubit is possibly strongly coupled to a
two-level system (TLS) modelling a representative mode of the environment,
giving rise to non-Markovian dynamics. In practice, the model can describe the
dynamics of two superconducting qubits in a LC circuit. The dissipation can for
example be described by a resistor~\cite{PhysRevB.94.184503} or by coupling one
of superconducting qubits to a lossy cavity~\cite{PhysRevLett.110.120501}. We
model the TLS and its interaction with the qubit by the following Hamiltonians,
\begin{align}
  \mathbf{H}_\text{tls} = -\frac{\omega_\text{tls}}{2}
  \boldsymbol{\sigma}_\text{tls}^z,
  \qquad
  \mathbf{H}_\text{int} = -J
  \boldsymbol{\sigma}_\text{q}^x\boldsymbol{\sigma}_\text{tls}^x,
\end{align}
where $\omega_\text{tls}$ is the frequency of the bath mode and $J$ the coupling
strength between the qubit and the TLS\@. The coupling of the TLS to the rest of
the environment is described by a standard Markovian master equation,
\begin{gather}
  \ic \frac{\text{d}}{\text{d}t} \boldsymbol{\rho}(t)
  =
  \left[\mathbf{H}(t), \boldsymbol{\rho}(t)\right]
  +
  \boldsymbol{\mathcal{L}}_{D} (\boldsymbol{\rho}),
  \\
  \boldsymbol{\mathcal{L}}_{D}(\boldsymbol{\rho})
  =
  \ic \kappa \sum\limits_{k = 1,2}
  \left(\mathbf{L}_k\boldsymbol{\rho} \mathbf{L}_k^\dagger -
    \frac{1}{2}\left\{\mathbf{L}_k^\dagger \mathbf{L}_k,
  \boldsymbol{\rho}\right\}\right),\nonumber
\end{gather}
where ${\mathbf{H}(t) = \mathbf{H}_\text{q}(t) + \mathbf{H}_\text{tls} +
\mathbf{H}_\text{int}}$ is the full Hamiltonian of the qubit and the TLS and
$\mathbf{L}_k$ the Lindblad operators. In what follows, we will refer to the two
parameters $J$ and $\kappa$ as coupling and rate, respectively, to point out
their different roles in the purification process, although formally they are of
the same nature. We assume the TLS and the bath to be initially in thermal
equilibrium characterized by
\begin{gather}
  \mathbf{L}_1 = \sqrt{N + 1} \boldsymbol{\sigma}_\text{tls}^-,
  \qquad
  \mathbf{L}_2 = \sqrt{N} \boldsymbol{\sigma}_\text{tls}^+,
\end{gather}
with $N = 1/(e^{\beta \omega_\text{tls}} - 1)$ and $\beta=k_B T$, $k_B$ and $T$
being, respectively, the Boltzmann constant and the temperature of the bath.
$\boldsymbol{\sigma}^{-}$ and $\boldsymbol{\sigma}^{+}$ are the standard
lowering and raising operators for two-level systems. The dynamics of the qubit
alone can be extracted as a partial trace over the TLS,
\begin{align}
  \boldsymbol{\rho}_\text{q} = \text{Tr}_\text{tls}(\boldsymbol{\rho}).
\end{align}
The density matrix of the joint system, i.e., qubit and TLS, is a $4\times 4$
Hermitian matrix which can be parameterized as
\begin{align}
  \boldsymbol{\rho}
  =
  \begin{pmatrix}
    x_1 & x_5 + \ic x_6 & x_7 + \ic x_8 & x_9 + \ic x_{10}\\
    x_5 - \ic x_6 & x_2 & x_{11} + \ic x_{12} & x_{13} + \ic x_{14}\\
    x_7 - \ic x_8 & x_{11} - \ic x_{12} & x_3 & x_{15} + \ic x_{16}\\
    x_9 - \ic x_{10} & x_{13} - \ic x_{14} & x_{15} - \ic x_{16} & x_4
  \end{pmatrix},
\end{align}
where the $x_i$ are real coefficients and $\sum_{i=1}^4 x_i=1$. The dynamical
space of the system therefore has 15 dimensions. After applying the rotating
wave approximation (RWA), see Appendix~\ref{app:rwa}, the dynamics can be
separated into four uncoupled subspaces. Only two of these contribute to the
qubit purity in which we are interested in, the other two are therefore
neglected. Technical details about the structure of the dynamical space are
given in Appendix~\ref{app:coord}. The definition of the subspaces is clarified
by introducing a new set of parameters:
\begin{gather}\label{eq:z-coord}
  \begin{aligned}
  z_1 &= x_1 + x_2 - 1/2,        &  z_5 &= x_7 + x_{13},\\
  z_2 &= x_{12},                 &  z_6 &= x_6 - x_{16},\\
  z_3 &= x_{11},                 &  z_7 &= x_8 + x_{14},\\
  z_4 &= -2 x_1 - x_2 - x_3,     &  z_8 &= x_5 - x_{15},
\end{aligned}
\end{gather}
in which the qubit purity reads
\begin{align}\label{eq:purity}
  P_\text{q} = \frac{1}{2} + 2\left( z_1^2 + z_5^2 + z_7^2\right).
\end{align}
We denote the subspaces associated with the coordinates $(z_1,z_2,z_3,z_4)$ and
$(z_5,z_6,z_7,z_8)$ by $S_1$ and $S_2$ . $S_1$ describes the
population of the qubit and its correlation with the TLS, while $S_2$ contains
information about the coherences of the qubit and the TLS. The equations of
motion on $S_1$ and $S_2$ are given by
\begin{align}\label{eq:dyn_s1}
  \begin{pmatrix}
    \dot{z}_1\\ \dot{z}_2\\ \dot{z}_3\\ \dot{z}_4
  \end{pmatrix}
  &= 2 J_1
  \begin{pmatrix}
    z_2\\ -z_1 - \frac{z_4 + 1}{2}\\ 0\\ 0
  \end{pmatrix}
  + 2 J_2
  \begin{pmatrix}
    z_3\\ 0\\ -z_1 - \frac{z_4 + 1}{2}\\ 0
  \end{pmatrix}\nonumber\\
  &\quad
  + 2 \alpha
  \begin{pmatrix}
    0\\ -z_3\\ z_2\\ 0
  \end{pmatrix}
  - \gamma
  \begin{pmatrix}
    0\\ \frac{z_2}{2} \\ \frac{z_3}{2} \\
    \frac{\gamma_1}{\gamma} + z_1 + z_4 + \frac{1}{2}
  \end{pmatrix},
\end{align}
and
\begin{align}\label{eq:dyn_s2}
  \begin{pmatrix}
    \dot{z}_5\\ \dot{z}_6\\ \dot{z}_7\\ \dot{z}_8
  \end{pmatrix}
  &= J_1
  \begin{pmatrix}
    z_6\\ -z_5 \\ -z_8\\ z_7
  \end{pmatrix}
  + J_2
  \begin{pmatrix}
    -z_8\\ z_7\\ -z_6 \\ z_5
  \end{pmatrix}\nonumber\\
  &\quad
  + 2 \alpha
  \begin{pmatrix}
    z_7\\ 0\\ -z_5\\ 0
  \end{pmatrix}
  - \frac{\gamma}{2}
  \begin{pmatrix}
    0\\ z_6\\ 0\\ z_8
  \end{pmatrix},
\end{align}
where we have introduced
\begin{equation}\label{notations}
  \begin{gathered}
    \begin{aligned}
    \delta(t) &= \omega_\text{q} + \varepsilon(t) - \omega_\text{tls}, \quad &
    \alpha(t) &= \frac{t}{2} \frac{\text{d}\delta}{\text{d}t},\\
    J_1 &= J\Cos{\delta t}, & J_2 &= J \Sin{\delta t},\\
    \gamma_1 &= \kappa (N + 1), & \gamma_2 &= \kappa N,\\
    \end{aligned}
    \ \\
    \text{with }
    \gamma = \gamma_1 + \gamma_2.
  \end{gathered}
\end{equation}
From a geometric point of view, $S_1$ is a 2-dimensional sphere in the space
$(z_1,z_2,z_3)$ defined by
\begin{align}\label{eq:r_s1}
  (z_1-c)^2+z_2^2+z_3^2
  =
  r(\gamma)
\end{align}
with its center $c=-(z_4+1)/2$ moving along the $z_1$-axis and radius $r$
decreasing with rate $\gamma$. $S_2$ describes a 3-dimensional sphere in the
space $(z_5,z_6,z_7,z_8)$ given by
\begin{align}\label{eq:r_s2}
  z_5^2+z_6^2+z_7^2+z_8^2
  =
  r'(\gamma)
\end{align}
with a fixed center at the origin and decreasing radius $r'$.

The initial state is constructed from the tensor product of the two separate
density matrices \cite{Basilewitsch2017c}
\begin{gather}\label{init_state}
  \begin{aligned}
  \boldsymbol{\rho} =& \boldsymbol{\rho}_\text{q} \otimes
  \boldsymbol{\rho}_\text{tls} + \boldsymbol{\rho}_\text{corr}\\
  =&
  \begin{pmatrix}
    a_\text{q} & \mu_\text{q} + \ic \nu_\text{q}\\
    \mu_\text{q} - \ic \nu_\text{q} & b_\text{q}
  \end{pmatrix}
  \otimes
  \begin{pmatrix}
    a_\text{tls} & 0\\
    0 & b_\text{tls},
  \end{pmatrix}\\
  &+
  \begin{pmatrix}
    0 & 0 & 0 & 0\\
    0 & 0 & \ic\xi & 0\\
    0 & -\ic\xi^* & 0 & 0\\
    0 & 0 & 0 & 0
  \end{pmatrix},
\end{aligned}
\end{gather}
where $a_k$ and $b_k$ are the ground- and exited state populations of qubit and
TLS in thermal equilibrium, which are defined by their respective energy level
splittings $\omega_{k}$ and the temperature. They can be expressed explicitly as
${a_k = \frac{e^{\beta\omega_k/2}}{2\ \cosh\left(\beta\omega_k /2\right)}}$ and
$b_k = 1 - a_k$. The parameters $\mu_\text{q}$ and $\nu_\text{q}$ are the coherences in the
reduced state of the qubit. We neglect coherences of the TLS assuming that it is
initially in a thermal state. Our analysis could also be carried out for a
non-thermal initial qubit population. Furthermore, we artificially add
coherences between the qubit and the TLS with the extra term
$\boldsymbol{\rho}_{\text{corr}}$. Since coherences $\xi$ give rise to
correlations between the qubit and the TLS, we refer to these coherences as
correlations throughout the paper.

If not stated otherwise, the parameters are set to ${\omega_\text{q} = 1,\
\omega_\text{tls} = 3,\ \beta = 1,\ J = 0.1}$ throughout as in
Ref.~\cite{Basilewitsch2017c}, allowing for a qualitative comparison of
the results, but in principle the parameters can be chosen arbitrarily. The only
constraint on the frequencies is $\omega_\text{q} < \omega_\text{tls}$ in order
for the qubit purity to be initially lower than the TLS purity. The coupling
strength $J$ obeys $J \ll \omega_\text{q}$ in order to satisfy the different
approximations made to establish the model system \cite{Basilewitsch2017c}.

\section{Purification of a qubit in a thermal state}\label{sec3}
In this section, we focus on the purification of a qubit in a thermal state.
This means, in particular, that the qubit has no initial coherence
(${\mu_\text{q} = \nu_\text{q} = 0}$) and all variables $z_{5}, \dots, z_{8}$
and their time derivatives vanish, see Eqs.~\eqref{eq:z-coord},
\eqref{eq:dyn_s1} and \eqref{init_state}. Therefore, we need to only consider the
dynamics in $S_1$, governed by Eq.~\eqref{eq:dyn_s1}, and neglect contributions
from $S_2$ for now. As a consequence, maximizing the purity $P_{\text{q}}$, see
Eq.~\eqref{eq:purity}, simplifies to maximizing $z_{1}$. In this case, using the
spherical symmetry, the dynamics can be further simplified by introducing
spherical coordinates,
\begin{equation}\label{eq:spher_coord}
  \begin{gathered}
    c = -\frac{z_4 + 1}{2},\\
    r \Sin{\theta} = z_1 - c,\\
    r \Cos{\theta}\Sin{\varphi} = z_2,\\
    r \Cos{\theta}\Cos{\varphi} = z_3.
  \end{gathered}
\end{equation}
Note that $r$ is identical with the one in Eq.~\eqref{eq:r_s1}. The full
dynamics of the qubit in these coordinates are then described by
\begin{subequations} \label{eqs1}
  \begin{align}
    \dot{r} &= -\frac{\gamma}{2} \left(r + (\eta - c)\Sin{\theta}\right),
    \\ \label{eq:s1:c}
    \dot{c} &= \frac{\gamma}{2} \left(r \Sin{\theta} + (\eta - c)\right),
    \\ \label{eq:s1:theta}
    \dot{\theta} &= -\frac{\gamma}{2}\frac{\eta - c}{r} \Cos{\theta} + 2 J
    \Cos{\delta t - \varphi},
    \\ \label{eq:s1:varphi}
    \dot{\varphi} &= 2\alpha - J \Tan{\theta}\Sin{\delta t - \varphi},
    \vphantom{\frac{\gamma}{2}}
  \end{align}
\end{subequations}
where $\eta = \gamma_1/\gamma - 1/2$ and the control field $\varepsilon(t)$ (see
Eq.~\eqref{eq:ham_q}) is present in the quantities $\delta(t)$ and
$\alpha(t)$.

Since we do not assume any initial coherence of the qubit, the qubit's purity
$P_{\text{q}}$ is completely determined by the dynamics on $S_1$. Using the
spherical coordinates of Eq.~\eqref{eq:spher_coord}, it can be expressed as
\begin{align} \label{eq:puritynew}
  P_\text{q} = \frac{1}{2} + 2(r \Sin{\theta} + c)^2.
\end{align}
Because $\varphi$ does not enter into the purity, we can define a new control,
\begin{align}
  u(t) = \delta t - \varphi.
\end{align}
Using Eq.~\eqref{eq:s1:varphi}, we arrive at
\begin{align}\label{eq:def_delta}
  \delta = \dot{u} - J \Tan{\theta} \Sin{u}.
\end{align}
This way we can first determine the optimal control strategy for $u(t)$ and
afterwards calculate the physical controls $\delta(t)$, respectively
$\varepsilon(t)$.

The North Pole of the $S_1$ sphere defined by $\theta = \pi/2$ is the state of
maximum purity, and we will denote its position on the $z_1$-axis by $Z
= r + c$. In principle, the maximum accessible purity can change over time since
the radius $r$ and the center $c$ of the sphere change. The time evolution of
$Z$ is governed by
\begin{align}\label{dotZ}
  \dot{Z} &= \dot{r} + \dot{c} = -\frac{\gamma}{2}(Z - \eta)\underbrace{(1 -
  \Sin{\theta})}_{\geq 0}.
\end{align}
Using Eqs.~\eqref{notations} and~\eqref{init_state}, it is straightforward to
show that ${\eta = a_\text{tls} - \frac{1}{2}}$. This quantity can be connected
to the initial TLS purity as $P_\text{tls}(0) = \frac{1}{2} + 2 \eta ^2$. The
behavior of $Z$ is different depending on whether qubit and TLS are initially
correlated or not. Hence, we examine both cases separately in the following.

\subsection{Time-optimal control in the correlation-free case}
If there is no initial correlation between qubit and TLS ($\xi = 0$), we
find the relation ${Z = a_\text{tls} - \frac{1}{2} = \eta}$ by evaluating the
initial state given by Eq.~\eqref{init_state} in terms of the coordinates of
Eqs.~\eqref{eq:z-coord} and \eqref{eq:spher_coord}. We therefore deduce from
Eq.~\eqref{dotZ} that $Z$, the North Pole of $S_{1}$, is a constant
of motion for correlation-free initial states. Moreover, this constant can be
used to simplify the differential system~\eqref{eqs1} even further by replacing
${c = Z - r = \eta - r}$. Effectively, the dynamics can then be described by
only two equations
\begin{subequations}
  \begin{align} \label{eqrtheta1}
    \dot{r} &= - \frac{\gamma}{2} r \left(1 + \Sin{\theta}\right),
    \\ \label{eqrtheta2}
    \dot{\theta} &= -\frac{\gamma}{2} \Cos{\theta} + 2J \Cos{u}.
  \end{align}
\end{subequations}
Without correlation ($\xi = 0$, implying $z_2 = z_3 = 0$), the initial state of
the system is the South Pole ($\theta = -\pi/2$) of $S_1$ as it can
be verified with Eq.~\eqref{eq:spher_coord}. Since $Z$ is a constant of motion,
the control strategy consists in performing a rotation to reach the North Pole
($\theta = \pi/2$) of the sphere as fast as possible. In the dissipation-free
case ($\gamma = 0$), the radius becomes constant and $\theta$ is rotating with
velocity $\dot{\theta} = 2 J \cos u$ (see Eq.~\eqref{eqrtheta2}). The
maximum speed for the rotation is reached with $u(t) = 0$ which corresponds to
the resonant case $\delta (t) = 0$ (see Eq.~\eqref{eq:def_delta}). This control
strategy does not change if the dissipation is taken into account. However, the
dissipative term slows down the rotation, which can be seen by the relative
opposite signs of the two terms in Eq.~\eqref{eqrtheta2}. Two scenarios can be
encountered according to the relative weights of the two terms, one in which the
dissipation dominates and a second where it can be viewed as a perturbation of
the unitary dynamics.

\begin{figure}[tb]
  \includegraphics[scale=1.0]{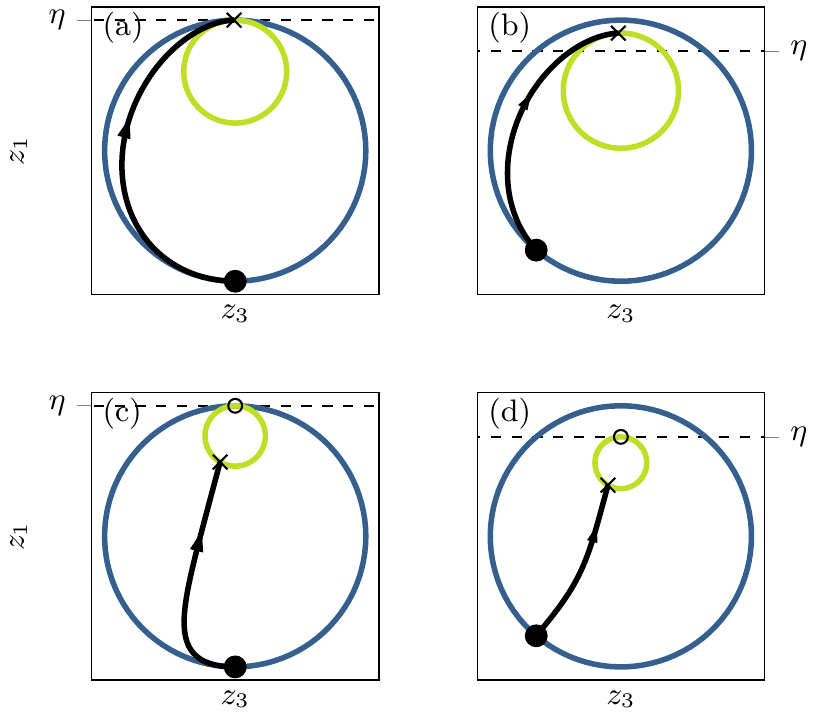}
  \caption{%
    (Color online) Optimal trajectories (in black) of the qubit in the
    $(z_1,z_3)$ - plane without(a) and with correlations (b) for
    non-Markovian dynamics. The initial and final states are represented,
    respectively, by a dot and a cross. In panels (c) and (d), the asymptotic
    steady states are indicated by circles. The blue (dark gray) and
    green (light gray) circles are the projections onto $(z_1,z_3)$ - plane of
    $S_1$ at the initial and final times. The amount of correlations
    added is equal to the maximum possible value $\xi = \xi_\text{max}$ (see
    Eq.~\eqref{eq:xi_max}) and $J = 4\,J_\text{min}$. Panel (c) shows the
    trajectory for the correlation-free Markovian case ($J=J_\text{min}/2$),
    while the correlated case is displayed in panel (d).
  }
  \label{fig:sphere}
\end{figure}

In general, we observe that the radius decreases exponentially while the
position $c$ of the center approaches asymptotically the value $\eta$. These
trajectories define the purity which can be reached by setting the position of
the north pole. On the other hand, the angular differential
equation~\eqref{eq:s1:theta} gives us information about the minimum time needed
to reach the state of maximum purity. For correlation-free initial states, the
angular equation (see Eq.~\eqref{eqrtheta2}) can be integrated analytically
leading to the minimum time \nolinebreak $T_{\textrm{min}}$, which is needed to
reach maximum purity on $S_1$,
\begin{align}\label{eq:tmin}
  T_{\textrm{min}} &= \int\limits_{-\pi/2}^{\pi/2}
  \frac{\text{d}\theta}{\dot{\theta}}
  =
  \frac{%
    8 \Arctan{\sqrt{\frac{4J + \gamma}{4J - \gamma}}}
  }{%
    \sqrt{(4J + \gamma)(4J - \gamma)}
  }.
\end{align}
In the zero dissipation limit $\gamma \rightarrow 0$, we recover the result
established in Ref.~\cite{Basilewitsch2017c} of ${T_{\textrm{min}}(\gamma
=0)=T_0 = \frac{\pi}{2J}}$. From Eq.~\eqref{eq:tmin} it can be seen that the
case $J \leq J_{\text{min}}$, with
\begin{align}\label{eq:Jmin}
  J_{\text{min}}=\gamma/4,
\end{align}
is not well defined. This scenario corresponds to the already mentioned case in
which the dissipation dominates, which can be attributed to the change
from non-Markovian to Markovian qubit dynamics. In the latter case, the
dissipative term becomes too large and a fixed point in $\theta$, i.e.,
$\dot{\theta}=0$, given by $\theta_\text{f} = \Arccos{4J/\gamma}$ arises. At the
fixed point, correlations between the qubit and TLS, which are build up during
the process, cannot be transformed into population anymore and therefore do not
further contribute to the purification. The North Pole is thus not accessible
and any gain in purification comes only from the exponential decrease in $r$
caused by the dissipation into the heat bath, see Fig.~\ref{fig:sphere}(c). This
is a remarkable feature, because naively the decrease of $r$ due to dissipation
would be connected to a loss of purity. Since the decrease in $r$ is maximized,
in this case, for $\theta=\theta_\text{f}$, the optimal strategy consists here
again in applying a zero control field $u(t)=0$. However, the final state cannot
be reached in finite time. Using a standard measure of non-Markovianity~\cite{lorenzo:2013}, 
we have also verified that the different
parameter regions can indeed be identified with the Markovian ($\gamma > 4J$)
and non-Markovian regimes ($\gamma < 4J$).

The trajectories for the non-Markovian and Markovian cases are plotted in
Fig.~\ref{fig:sphere}(a) and (c). Figure~\ref{gamma-T} displays the dependence
of the minimum time on the ratio $\gamma/J$ for correlated and uncorrelated
initial states. The sharp transition to the Markovian regime can be observed at
$\gamma = 4 J$, indicated by the divergence of the purification time.
Figure~\ref{gamma-T} shows that the purification time for correlated initial
states is lower than for uncorrelated ones. As can be seen in
Fig.~\ref{fig:sphere}(b), this is a consequence of the position of the initial
state which is closer to the equator of $S_1$.


\subsection{Time-optimal control with correlated states}
Adding $\xi$ - correlations between qubit and TLS to the initial
state~\eqref{init_state} changes the dynamics because $Z(0) \neq \eta$ and
therefore $Z$ is not constant anymore, as shown in Eq.~\eqref{dotZ}. Although it
makes a difference whether $\xi$ is real or imaginary, we will only consider
real $\xi$ in what follows. This is because a purely imaginary $\xi$ only
modifies the initial value of $\varphi$. The control field can always be chosen
so that it produces a short and strong $\alpha$-pulse in order to rotate
$\varphi$ to $0$, see Eq.~\eqref{eq:s1:varphi} and
Ref.~\cite{Basilewitsch2017c}. Since this rotation can be made arbitrarily fast
(at least theoretically), we focus on the time-optimal solution for the
remaining control problem which coincides with the case of initially real $\xi$.

Arbitrarily large correlations cannot be introduced due to the physical
constraint of the density matrix being positive semi-definite. An eigenvalue
analysis reveals that the maximum amount of correlation is
\begin{gather}\label{eq:xi_max}
  \xi_\text{max} = \sqrt{a_\text{q}a_\text{tls}b_\text{q} b_\text{tls}}.
\end{gather}
The dynamics of the maximal reachable purity depends on the initial value
of $Z - \eta$. Using the definition of the initial state~\eqref{init_state}, we
find
\begin{align}\label{Z0-eta}
  Z(0) -\eta = \sqrt{\left(\frac{a_\text{tls} - a_\text{q}}{2}\right)^2 + \xi^2}
  - \frac{a_\text{tls} - a_\text{q}}{2} \geq 0,
\end{align}
from which together with Eq.~\eqref{dotZ}, we can conclude that ${\dot{Z} \leq
0}$ and
\begin{equation}\label{eq:Z-eta}
  Z(t)-\eta
  =
  (Z(0)-\eta)\exp\left[%
    -\frac{\gamma}{2}\int_0^t(1-\Sin{\theta})\text{d}t'
  \right].
\end{equation}
Correlations therefore increase the initially accessible purity which then
decays asymptotically to $\eta$, the same value as in the uncorrelated case.
This decay is caused by the decrease of the radius $r$, which can be written as
\begin{equation}
  \dot{r}
  =
  -\frac{\gamma}{2}[\underbrace{Z-c}_{=r}+(\eta-c)\Sin{\theta}].
\end{equation}
To prove that $r$ is monotonically decreasing, we distinguish two
cases.
\begin{itemize}
  \item $\eta - c \geq 0$:\\
    In this situation, together with Eq.~\eqref{eq:Z-eta}, we can estimate
    \begin{align}
      \dot{r} \leq -\frac{\gamma}{2} \left( \eta - c \right)
      \left( 1 + \Sin{\theta} \right) \leq 0.
    \end{align}
  \item $\eta - c \leq 0$:\\
    From Eq.~\eqref{eq:s1:c}, we can deduce the maximum of $c$ during the process
    as
    \begin{align}
      c_\text{max} = r \Sin{\theta} + \eta \leq r + \eta.
    \end{align}
    Using this relation, an upper limit for $\dot{r}$ is given by:
    \begin{align}
      \dot{r} = -\frac{\gamma}{2} \left( r + \underbrace{\left( \eta - c \right)
      \Sin{\theta}}_{\geq -r}\right) \leq 0.
  \end{align}
\end{itemize}

\begin{figure}[tb]
  \centering
  \includegraphics[scale=1.0]{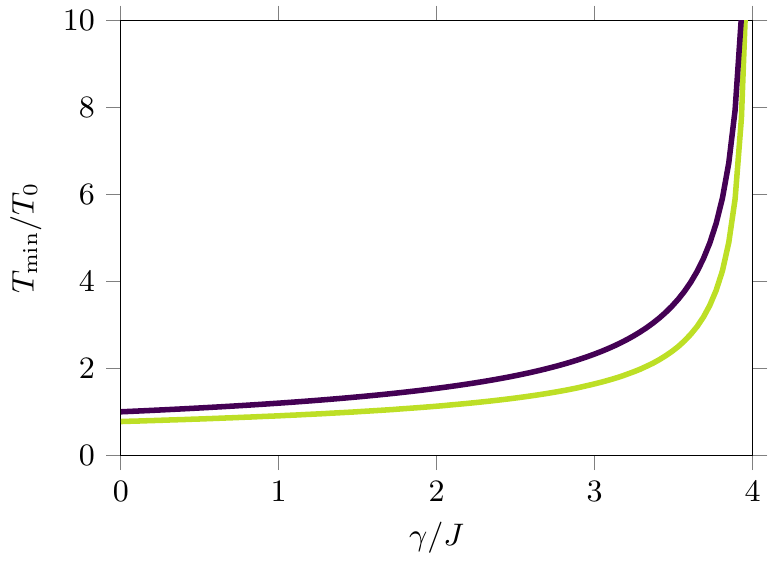}
  \caption{%
    (Color online) Normalized minimum time $T_{\textrm{min}}/T_0$ to reach the
    north pole of $S_1$ as a function of $\gamma/J$ for the correlated (green or
    light gray) and uncorrelated (purple or dark gray) initial states.
    $T_0$ is the purification time for $\gamma = 0$ and the parameter
    $\xi$ is set to $\xi_\text{max}$ in the correlated case.
  }
  \label{gamma-T}
\end{figure}

As before, we study the time needed to reach the state of maximum purity by
examining the angular dynamics which are governed by (see
Eq.~\eqref{eq:s1:theta})
\begin{equation}\label{eqnew}
  \dot{\theta}
  = \frac{\gamma}{2} \frac{Z-\eta}{r} \Cos{\theta} - \frac{\gamma}{2}
  \Cos{\theta} + 2J\Cos{u}.
\end{equation}
Equation~\eqref{eqnew} is similar to the correlation-free
version~\eqref{eqrtheta2} but added by a new term, which is always positive in
the region of interest $\theta \in [-\pi/2,\pi/2]$. As before, the latter
driving term is strongest for $u(t)=0$. The purification time is lower than in
the uncorrelated case due to the additional first positive term, which increases
the effective driving speed. In addition, correlations change the initial state
for $\theta$ given by $\theta(0)=\arccos\left(\xi/r(0)\right)$, which leads to
a shorter distance towards the $S_{1}$ north pole to be covered. In particular,
the minimum time for $\gamma=0$ is $T_0=\frac{\pi/2-\theta (0)}{2J}$. The
angular dynamics~\eqref{eqnew} cannot be integrated analytically anymore, but
Fig.~\ref{gamma-T} shows the numerically calculated times in comparison to the
analytical results in the correlation-free case. Interestingly the same
divergence for $\gamma > 4J$, which corresponds to the transition between
Markovian and non-Markovian behavior, can be observed. Physically, this means
that if the dissipation becomes too large in comparison with the coupling $J$, the
dynamics become Markovian and purification takes an infinite amount
of time. The optimal trajectory for the Markovian case is plotted in
Fig.~\ref{fig:sphere}(d).  Nevertheless, we observe that the final state has
a lower purity than the initial North Pole even in the case of non-Markovian
dynamics. This is due to the decrease of $Z$ over time. However, the final
purity is still higher than in the correlation-free case, i.e., with $\xi=0$.
The optimal trajectory for this situation is shown in Fig.~\ref{fig:sphere}(b).

\subsection{Role of initial correlations for the existence of a fixed point}
Despite being able to reach higher purity in a shorter time, the non-Markovian
regime has the drawback the state of maximal purity not being stable.
Therefore, after reaching the target state, qubit and TLS have to be
decoupled or the purity of the qubit will decrease. This is not the case for
Markovian dynamics as shown in Fig.~\ref{fig:sphere}(c). The angular fixed point
is reached and the system tends continuously to the state of maximum purity,
which is in return never reached exactly in finite time.

\begin{figure}[tb]
  \centering
  \includegraphics[scale=1.0]{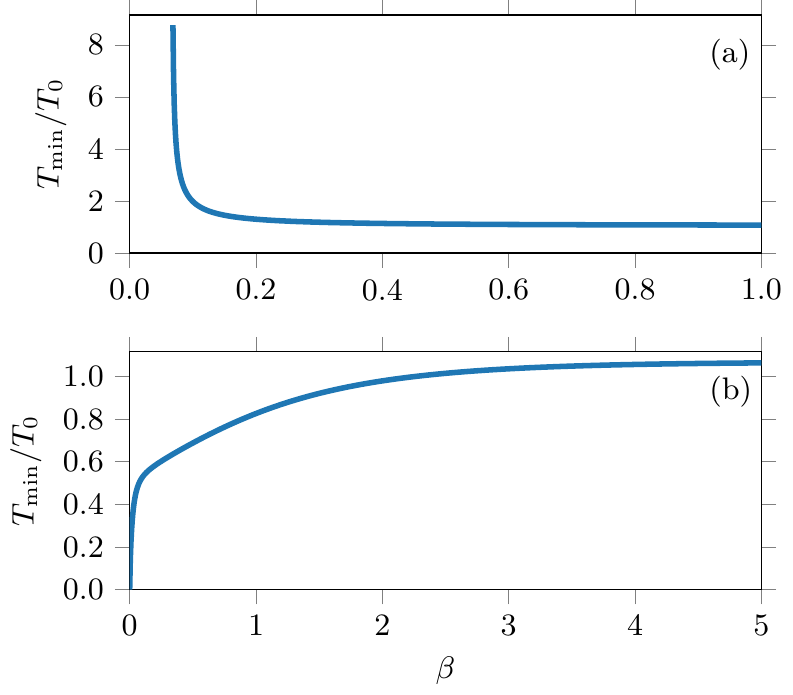}
  \caption{%
    (Color online) Purification time as a function of the temperature. Panel (a)
    represents the correlation-free initial state in which we observe the same
    behavior as in Fig.~\ref{gamma-T}. For a specific threshold, the
    purification time diverges. If correlations are added (panel (b)) then the
    angular fixed point is resolved and the target state can be reached in
    finite time. $\xi = \xi_\text{max}$ fixed, but its value depends on $\beta$.
    The parameter $T_0$ was chosen as $T_0(\xi=0)$ in panel (b) for the sake of
    comparison with panel (a).
  }
  \label{fig:beta-T}
\end{figure}

Figure~\ref{fig:beta-T} displays the dependence of the purification on the
inverse temperature for uncorrelated and correlated initial
states. For large $\beta$, we approach the same purification time in the
two situations because the amount of allowed correlations goes to zero in this
limit (see Eq.~\eqref{eq:xi_max}). The minimum time $T_{\textrm{min}}$ is
slightly larger than $T_0$ due the dissipation terms which are different from
zero even at low temperature (see Eq.~\eqref{notations}),
\begin{align}
    \lim\limits_{\beta \rightarrow \infty}{\gamma} = \kappa \neq 0.
\end{align}
Surprisingly, the dynamics have a different behavior for small $\beta$. In
Fig.~\ref{fig:beta-T}(a), the transition to the Markovian regime is similar to
the one of Fig.~\ref{gamma-T} with a divergence of the purification time. For
correlated initial states (Fig.~\ref{fig:beta-T}(b)), we observe that for low
temperatures, i.e.\ large $\beta$, the purification time decreases and
approaches zero. This suggests that the angular fixed point can be resolved by
adding correlations, as described below.

The following discussion describes the behaviour of ${\dot{\theta} = 0}$. We
will refer to the value of $\theta$ at which its derivative vanishes as
angular fixed point $\theta_\text{f}$, although it is not a fixed point of the
full dynamics i.e. a steady state.

For correlated initial states, the fixed point equation reads
\begin{equation}\label{eqfixed}
  \theta_\text{f} = \Arccos{\frac{4 J}{\gamma}\frac{r}{\eta - c}}
  =
  \Arccos{\frac{4 J}{\gamma}\frac{Z - c}{\eta - c}}.
\end{equation}
Note that the value of $\theta_\text{f}$ depends on $r$ and $c$ and therefore
can change over time. It is only a fixed point in the sense that if ${\theta =
\theta_\text{f}}$ is reached, it will not change its value anymore, even
though $r$ and $c$ will still continue to vary.

This fixed point is only defined if $\frac{4 J}{\gamma} \frac{r}{\eta - c} > 1$,
otherwise there is no solution to Eq.~\eqref{eqfixed} and no fixed point occurs.
Recall that, for uncorrelated initial states, we found ${Z = \eta}$ and therefore
we recover $J \leq \frac{\gamma}{4}$ as the condition for the existence of the
fixed point. In general, as can be seen from Eq.~\eqref{Z0-eta},
the second term is always larger or equal to one and $J > \frac{\gamma}{4}$
leads to fixed point-free dynamics. However, if initial correlations between
qubit and TLS are introduced, the fixed point can be resolved for $J <
\frac{\gamma}{4}$. From Eq.~\eqref{eqfixed}, we can calculate the maximum amount
of correlations for which the fixed point is still defined,
\begin{equation}\label{eq:theta_fixed}
  \xi_\text{fixed} = \pm \frac{a_\text{tls} - a_\text{q}}{2}\sqrt{1 -
  \left(\frac{\gamma}{4 J}\right)^2}.
\end{equation}
If more correlations are included, there is no fixed point present
initially. Nevertheless, a fixed point, into which the dynamics may eventually
run, can still occur during the time evolution itself.

\begin{figure}[tb]
  \centering
  \includegraphics[scale=1.0]{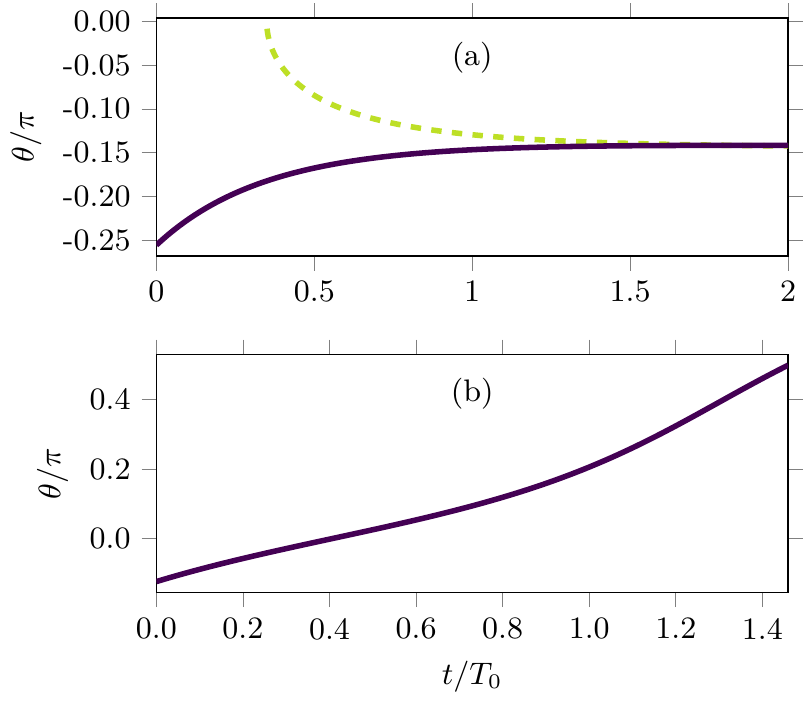}
  \caption{%
    (Color online) Time evolution (solid line) of the angle $\theta$ in the
    Markovian regime $J = 0.9\,J_\text{min}$ (cf. Eq. \eqref{eq:Jmin}). The
    correlations are set to $\xi = 2\,\xi_\text{fixed}$ and $\xi = 5
    \,\xi_\text{fixed}$ in panels (a) and (b), respectively. The dashed green
    line in panel (a) depicts the position of the fixed point $\theta_\text{f}$.
}
\label{t-theta}
\end{figure}

Figure~\ref{t-theta} displays the dynamics of $\theta$ and the time
evolution of the value of $\theta_\text{f}$. It can be seen that exceeding the
preceding bound~\eqref{eq:theta_fixed} even further (i.e.\ comparing
Fig.~\ref{t-theta}(a) and (b)) prevents the fixed point from arising also during
the time evolution. The system can reach the angle $\theta = \pi/2$ and
therefore the state of maximum purity in finite time. In contrast to
correlation-free initial states, this conclusion is true for any temperature
with sufficient initial correlations. Note that the limiting boundary for a
valid density matrix has to be satisfied.

\begin{figure}[tb]
  \centering
  \includegraphics[scale=1.0]{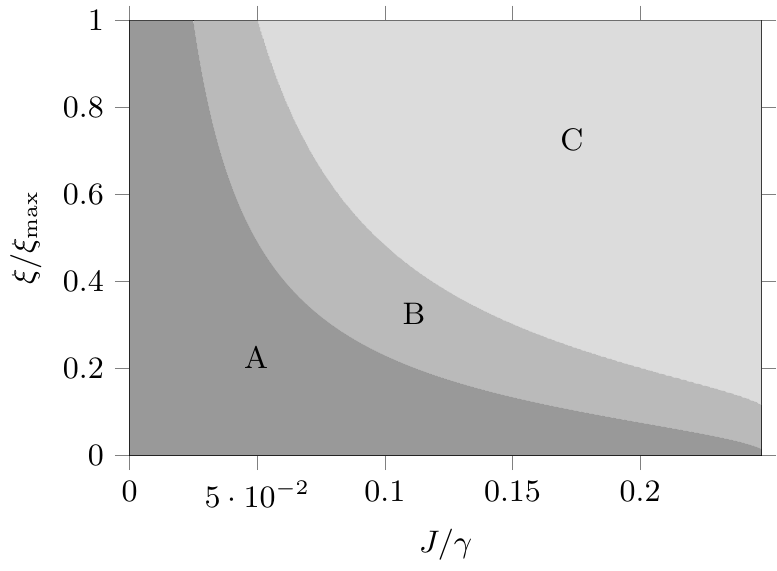}
  \caption{%
    Existence of an angular fixed point in $\theta$ as a function of correlation
    and coupling strength ($\beta = 0.1$). In region A, a fixed point is
    initially defined, while in region B there no fixed point accessible
    initially, but during the time evolution of the system. Area C corresponds
    to the parameter space in which no fixed point occurs in the time in which
    the final state is reached.
  }
\label{fixed_point}
\end{figure}

In general, the dynamics of the system can be split into different regimes,
depending on the correlations, which are shown in Fig.~\ref{fixed_point}. The
different zones describe the regime in which no fixed point is present
initially, the case where the fixed point arises during the evolution and the
region in which no fixed point occurs during the whole purification process.
Interestingly, we can go from one regime to the other by controlling the amount
of correlations between the qubit and the TLS\@. Although it is possible to
purify the system in region C in finite time, as it is in the non-Markovian
regime, it is important to point out that non-Markovianity is a feature of the
dynamical map, which does not depend on the initial state~\cite{dajka:2011,
smirne:2010, wissmann:2013}.

\section{Role of initial coherences and correlations for the control strategy}\label{sec4}
We investigate in this section the joint influence of correlations in the
presence of initial qubit coherences, i.e., $\mu_\text{q}, \nu_\text{q} \neq 0$,
on the optimal strategy designed in Sec.~\ref{sec3}. In particular, qubit
coherences lead to a dynamics on $S_2$ since $z_{5}$ or $z_{7}$ are not
vanishing anymore, see Eq.~\eqref{eq:z-coord} and~\eqref{eq:dyn_s2}. Hence, the
qubit purity $P_{\text{q}}$ gets now simultaneous contributions from both
spheres. Using Eq.~\eqref{eq:purity}, the purity can be split into two different
contributions. The terms proportional to $z_1^2$ will be called contribution
from $S_1$ and the other terms will be assigned to $S_2$.

\begin{figure}[tb]
  \centering
  \includegraphics[scale=1.0]{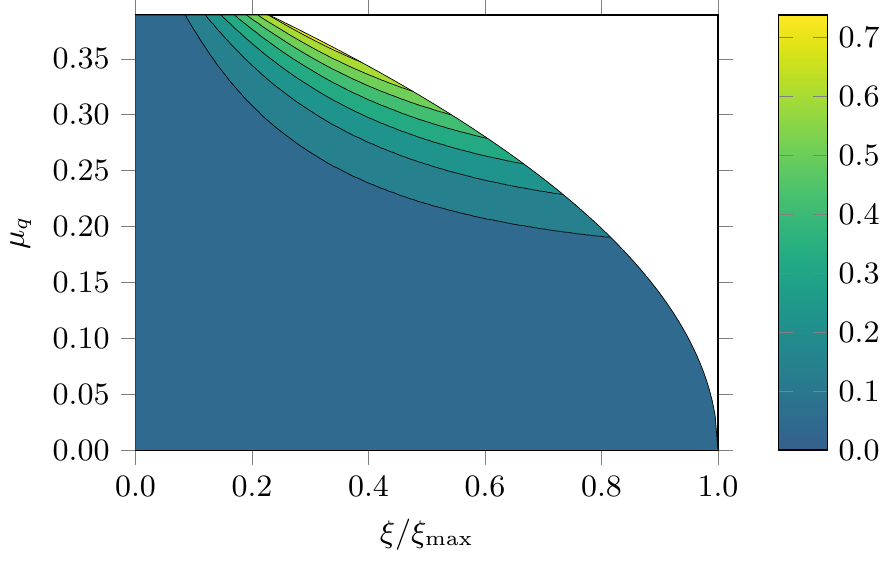}
  \caption{%
    (Color online) Parameter space of the correlations $\xi$ and coherences
    $\mu_\text{q}$ for which the density matrix is defined. The color code
    indicates the relative difference (Eq.~\eqref{eq:P_diff}) between the
    maximum purity reached during the evolution and the purity at ${\theta
    = \pi/2}$ in per cent.
  }
  \label{fig:limit_xi_mu}
\end{figure}

Therefore, it is interesting to study whether the dynamics on $S_2$ change the
procedure to get the highest overall purity $P_{\text{q}}$. For the resonant
case, $\delta = \alpha = 0$, the equations on $S_2$ are
\begin{subequations}\label{eqs2}
  \begin{align}
    \dot{z}_5 &= J z_6,\\
    \dot{z}_6 &= -J z_5 - \frac{\gamma}{2} z_6,\\
    \dot{z}_7 &= -J z_8,\\
    \dot{z}_8 &= J z_7 - \frac{\gamma}{2} z_8,
  \end{align}
\end{subequations}
with the initial conditions
\begin{align}
  z_5(0) &= \mu_\text{q}, & z_7(0) &= \nu_\text{q},\nonumber\\
  z_6(0) &= 0, & z_8(0) &= 0.
\end{align}
Because the equations for $z_5$ and $z_7$ are decoupled and only their squared
sum enters into the purity, it is sufficient to consider $z_7 = 0$ or
equivalently only real coherences. The equations of motion are identical to the
ones of a damped harmonic oscillator. The solution reads
\begin{align}\label{eq:z5}
  z_5(t) = \mu_\text{q} \sqrt{1 + \left(\frac{\gamma}{4 \omega}\right)^2}
  \Cos{\omega t - \Arctan{\frac{\gamma}{4\omega}}}
  \Exp{-\frac{\gamma}{4} t},
\end{align}
with $\omega = \sqrt{J^2 - \gamma^2/16}$. This describes an oscillating
behavior damped by an exponential decay having its maximum at $t = 0$. The
purity contribution from $S_{2}$ is therefore maximal in the initial state. As
in Sec.~\ref{sec3}, we can identify the Markovian limit $\gamma \geq 4J$ in
which the cosine function turns into a hyperbolic cosine and $z_5$ is
monotonically decreasing. We focus below only on the non-Markovian case. Caution
has to be made on the allowed range of parameters $\mu_\text{q}$ and $\xi$. For
vanishing coherences, the maximum value of $\xi$ has already been calculated in
Eq.~\eqref{eq:xi_max} and this computation can be done for
$\mu_\text{q}$ in a similar way. If both coherences and correlations are present
then the limits are determined numerically. We compute the maximum value of
$\mu_\text{q}$ for which the density matrix for a given $\xi$ has non-negative
eigenvalues. The allowed parameter region is plotted in
Fig.~\ref{fig:limit_xi_mu}.

At this point, we already know how to maximise the purity contributions
from $S_1$ and $S_2$ separately. It is however not clear how the overall purity
behaves. We again consider the cases of correlated and uncorrelated initial
states separately.

\begin{figure}[tb]
  \centering
  \includegraphics[scale=1.0]{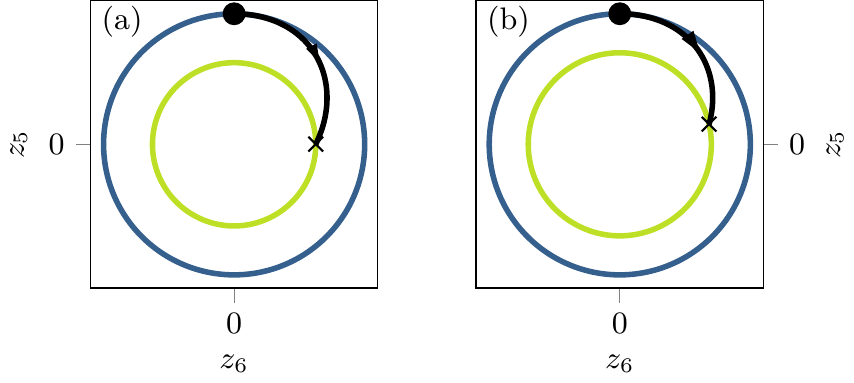}
  \caption{%
    (Color online) Optimal trajectories on $S_2$ in the $(z_5,z_6)$- plane
    without (a) and with (b) initial correlations. The initial
    and final points are represented, respectively, by a dot and a cross. The
    control time is set to $T_{\text{min}}$, see Eq.~\eqref{eq:tmin}.
    Parameters are set to $\xi=\xi_\text{max}/2$ and $\mu_\text{q}
    = \mu_\text{q, max}$. Note that $\mu_\text{q, max}$ depends on $\xi$. The
    blue (dark gray) and green (light gray) circles are the projections of $S_2$
    onto the $(z_5,z_6)$- plane at initial and final times.
  }
  \label{fig:sphere_s2}
\end{figure}

\begin{figure}[bt]
  \centering
  \includegraphics[scale=1.0]{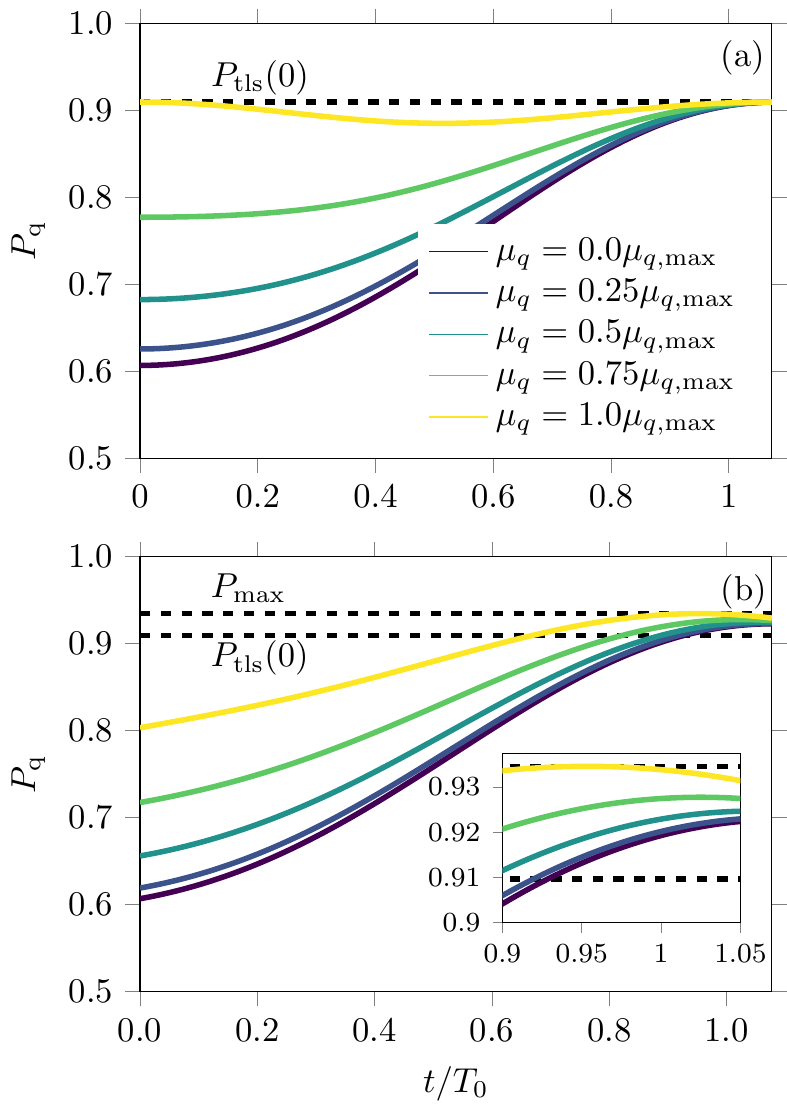}
  \caption{%
    (Color online) Time evolution of the qubit purity $P_\text{q}$ of the qubit
    for uncorrelated (a) and correlated (b) initial states with different
    coherences $\mu_\text{q} \in [0,\mu_\text{q,max}]$, where $\mu_\text{q,max}$
    is the maximum allowed value of the coherence. The parameter $\xi$ is set to
    $0$ and $\xi_\text{max}/2$ in (a) and (b) respectively. The horizontal
    dashed lines depict the initial purity of the TLS, $P_\text{tls}(0)$ and the
    maximum value $P_{\textrm{max}}$ of $P_\text{q}$.
  }
  \label{fig:purity}
\end{figure}

In the uncorrelated situation, we combine Eqs.~\eqref{eq:tmin} and \eqref{eq:z5}
to observe that, at time $T_{\textrm{min}}$, where the purity is maximum on
$S_1$, the contribution of $S_2$ vanishes i.e. $z_5(T_{\textrm{min}}) = 0$. The
corresponding trajectory is plotted in Fig.~\ref{fig:sphere_s2}(a). As shown in
Fig.~\ref{fig:purity}(a), numerical simulations reveal that the dynamics
on $S_2$ are not relevant at all since the maximum purity and the time to reach
it are the same as the ones on $S_1$ for any value of $\mu_\text{q}$. In particular,
the best final purity is limited by the initial purity of the TLS.

However, this behavior changes if correlations are considered. As can be seen in
Eq.~\eqref{eqs2}, they do not affect the dynamics on $S_2$ but they reduce the
time needed to reach the north pole on $S_1$ and therefore introduce a phase
shift between $S_1$ and $S_2$. Due to this shorter time, the contribution from
$S_2$ has not completely vanished yet and the overall purity can increase, see
Fig.~\ref{fig:sphere_s2}(b). This maximum amount of purity, which is reached
during the purification process, is called $P_\text{max}$. The color code in
Fig.~\ref{fig:limit_xi_mu} indicates
\begin{align}\label{eq:P_diff}
  \Delta P = \frac{P_\text{max}}{P(T_\text{min})} - 1.
\end{align}
This corresponds to the relative purity which is gained by taking into account
the combined dynamics of $S_1$ and $S_2$. The numerical results of
Fig.~\ref{fig:purity}(b) demonstrate that, in case of initial correlations,
qubit coherences can be transformed into an additional gain of population and
therefore break the limit of the TLS purity. Note that this is not possible with
correlation-free initial states as shown in Fig.~\ref{fig:sphere_s2}(a)
and~\ref{fig:purity}(a). Moreover, it can be seen in
Fig.~\ref{fig:purity}(b) that, while the maximally accessible purity increases,
the minimum time needed to reach it decreases as coherences increase. In other
words, qubit coherences improve both total time and final purity of the control
scheme, but require qubit and TLS to be initially correlated.

\section{Conclusions}\label{sec5}
We have investigated control of a qubit coupled to a structured reservoir, which
is composed of a well-defined and strongly coupled mode and a thermal bath.
This model can be realized experimentally with superconducting qubits. We assume that only the energy splitting
of the qubit can be changed by the external control field. Using a geometric
description of the control problem, we show that the time-optimal protocol to
purify the qubit is based on fulfilling a resonance condition between the qubit
and the reservoir mode. This result is valid for any coupling strength between
the qubit and the environment and for any decay rate of the thermal bath.
Non-Markovianity of the qubit dynamics does not modify the control strategy, but
reduces the time to reach the state of maximum purity.  Introducing
strong correlations between qubit and TLS accelerates the process even further.
The role of initial qubit coherences has been investigated as well:
Combined correlations and qubit coherences speed up the control process
and improve the final purity of the qubit even more.

Our study and the possibility to describe geometrically purification of a qubit
in contact with a structured reservoir pave the way to future investigations. In
particular, it would be interesting to generalize this model to more complex
control scenarios in which the external field can be applied also in other
directions. For instance, the qubit coherence can be modified by a $\sigma_x$-
control, which could be combined with the $\sigma_z$- control used here to
enhance or speed up the purification process.  Another intriguing avenue is the
application of this approach to algorithmic cooling (AC) in which similar model
systems are considered~\cite{AC1, AC2, AC3}. To the best of our knowledge, AC
methods neglect the interaction between the bath and the qubits during the first
step of the cooling, i.e.\ the entropy exchange. This approximation could be
avoided by generalizing the results of this work to the case of $n$ qubits
($n>1$) and $m$ reset qubits or modes ($m\geq 1$).

\appendix

\section{Rotating wave approximation}
\label{app:rwa}
Starting from the Hamiltonian governing the dynamics of the model system,
\begin{align}
  \mathbf{H}(t) = \underbrace{- \frac{\omega_\text{q} + \varepsilon(t)}{2}\
  \boldsymbol{\sigma}_\text{q}^z - \frac{\omega_\text{tls}}{2} \
  \boldsymbol{\sigma}_\text{tls}^z}_{\mathbf{H}_0}
  - J \boldsymbol{\sigma}_\text{q}^x \boldsymbol{\sigma}_\text{tls}^x,
\end{align}
we can transform the Hamiltonian using the unitary transformation $\mathbf{U}(t)
= \Exp{\ic \mathbf{H}_0 t}$ to get
\begin{widetext}
  \begin{align}
    \mathbf{H}'(t) &= \mathbf{U}(t) \mathbf{H}(t) \mathbf{U}^\dagger(t) - \ic\
    \mathbf{U}(t) \frac{\text{d}\mathbf{U}^\dagger(t)}{\text{d}t}\nonumber\\
    &=
    \begin{pmatrix}
      \frac{t}{2}\Der{\varepsilon(t)}{t} & 0 & 0 & -J \Exp{-\ic t (\omega_\text{q} +
      \omega_\text{tls} + \varepsilon(t))}\\
      0 & \frac{t}{2}\Der{\varepsilon(t)}{t} & -J \Exp{-\ic t (\omega_\text{q} -
      \omega_\text{tls} + \varepsilon(t))} & 0\\
      0 & -J \Exp{\ic t (\omega_\text{q} - \omega_\text{tls} + \varepsilon(t))} &
      -\frac{t}{2}\Der{\varepsilon(t)}{t} & 0\\
      -J \Exp{\ic t (\omega_\text{q} + \omega_\text{tls} + \varepsilon(t))} & 0 & 0 &
      -\frac{t}{2}\Der{\varepsilon(t)}{t}
    \end{pmatrix}
  \end{align}
  The terms $J \Exp{\pm \ic t (\omega_\text{q} + \omega_\text{tls} + \varepsilon(t))}$
  are oscillating fast and average to zero on a short timescale.
  Therefore we neglect these terms and obtain the Hamiltonian after
  the rotating wave approximation (RWA) as
  \begin{align}\label{hamiltonian_rwa}
    \mathbf{H}^\text{rwa}(t) &=
    \begin{pmatrix}
      \frac{t}{2}\Der{\varepsilon(t)}{t} & 0 & 0 & 0\\
      0 & \frac{t}{2}\Der{\varepsilon(t)}{t} & -J \Exp{-\ic t (\omega_\text{q} -
      \omega_\text{tls} + \varepsilon(t))} & 0\\
      0 & -J \Exp{\ic t (\omega_\text{q} - \omega_\text{tls} + \varepsilon(t))} &
      -\frac{t}{2}\Der{\varepsilon(t)}{t} & 0\\
      0 & 0 & 0 & -\frac{t}{2}\Der{\varepsilon(t)}{t}
    \end{pmatrix}.
  \end{align}
  The Lindblad operators $\mathbf{L}_{1/2} = \sqrt{\gamma_{1/2}}
  \boldsymbol{\sigma}_\text{tls}^\pm$ have to be transformed in the same manner
  resulting in $\mathbf{L}_{1/2}^\text{rwa} = \Exp{-\ic \omega_\text{tls} t}
  \mathbf{L}_{1/2}$. We can analyze the dynamics of the transformed system using
  the Lindblad equation
  \begin{align}
    \Der{}{t} \bm{\rho}^\text{rwa}(t) = -\ic \left[\mathbf{H}^\text{rwa}(t),
    \bm{\rho}^\text{rwa}\right] + \sum\limits_{k = 1,2} \left(
      \mathbf{L}_k^\text{rwa} \bm{\rho}^\text{rwa}
      {\mathbf{L}_k^\text{rwa}}^\dagger - \frac{1}{2} \left\{
        {\mathbf{L}_k^\text{rwa}}^\dagger \mathbf{L}_k^\text{rwa},
    \bm{\rho}^\text{rwa}\right\}\right)
  \end{align}
\end{widetext}

\section{Coordinate transformation}
\label{app:coord}
We consider a Hamiltonian of the form \eqref{hamiltonian_rwa}
\begin{align}
  \mathbf{H} =
  \begin{pmatrix}
    \alpha & 0 & 0 & 0\\
    0 & \alpha & -J \Exp{-\ic \delta t} & 0\\
    0 & -J \Exp{\ic \delta t} & -\alpha & 0\\
    0 & 0 & 0 & -\alpha
  \end{pmatrix},
\end{align}
and parameterize the full density matrix as
\begin{align}
  \boldsymbol{\rho} =
  \begin{pmatrix}
    x_1 & x_5 + \ic x_6 & x_7 + \ic x_8 & x_9 + \ic x_{10}\\
    x_5 - \ic x_6 & x_2 & x_{11} + \ic x_{12} & x_{13} + \ic x_{14}\\
    x_7 - \ic x_8 & x_{11} - \ic x_{12} & x_3 & x_{15} + \ic x_{16}\\
    x_9 - \ic x_{10} & x_{13} - \ic x_{14} & x_{15} - \ic x_{16} & x_4
  \end{pmatrix}.
\end{align}
Then the Markovian master equation
\begin{align}
  \ic \frac{\text{d}}{\text{d}t} \boldsymbol{\rho}(t) =& \left[\mathbf{H}(t),
  \boldsymbol{\rho}(t)\right] \nonumber\\
  &+ \ic \sum\limits_{k = 1,2}
  \left(\mathbf{L}_k\boldsymbol{\rho} \mathbf{L}_k^\dagger -
    \frac{1}{2}\left\{\mathbf{L}_k^\dagger \mathbf{L}_k,
  \boldsymbol{\rho}\right\}\right),
\end{align}
with the Lindblad operators
\begin{equation}
  \begin{cases}
    \mathbf{L}_1 = \sqrt{\gamma_1} \boldsymbol{\sigma}_\text{tls}^-,\\
    \mathbf{L}_2 = \sqrt{\gamma_2} \boldsymbol{\sigma}_\text{tls}^+,\\
  \end{cases}
\end{equation}
gives us a set of differential equations for the parameters $\pmb{x} =
(x_1,\dots,x_{16})$.
\begin{widetext}
These equations can be written in the form
\begin{align}
  \dot{\pmb{x}} = \mathbf{f}_0(\pmb{x}) + J_1\ \mathbf{f}_1(\pmb{x}) +
  J_2\ \mathbf{f}_2(\pmb{x}) + \alpha \ \mathbf{f}_3(\pmb{x}),
\end{align}
with $J_1 = J\Cos{\delta t},\ J_2 = J \Sin{\delta t}$ and
\begin{align}
  \mathbf{f}_0(\pmb{x}) =
    \gamma_1 \begin{pmatrix}
    x_2\\-x_2\\x_4\\-x_4\\-x_5/2\\-x_6/2\\x_{13}\\x_{14}\\-x_9/2\\-x_{10}/2\\-x_{11}/2\\-x_{12}/2\\-x_{13}\\-x_{14}\\-x_{15}/2\\-x_{16}/2
    \end{pmatrix}
    +
    \gamma_2 \begin{pmatrix}
    -x_1\\x_1\\-x_3\\x_3\\-x_5/2\\-x_6/2\\-x_7\\-x_8\\-x_9/2\\-x_{10}/2\\-x_{11}/2\\-x_{12}/2\\-x_7\\-x_8\\-x_{15}/2\\-x_{16}/2
  \end{pmatrix},\
  \mathbf{f}_1(\pmb{x}) =
    \begin{pmatrix}0\\2x_{12}\\-2x_{12}\\0\\x_8\\-x_7\\x_6\\-x_5\\0\\0\\0\\x_3-x_2\\-x_{16}\\x_{15}\\-x_{14}\\x_{13}\end{pmatrix},\
   \mathbf{f}_2(\pmb{x}) =
    \begin{pmatrix}0\\2x_{11}\\-2x_{11}\\0\\x_7\\x_8\\-x_5\\-x_6\\0\\0\\x_3-x_2\\0\\x_{15}\\x_{16}\\-x_{13}\\-x_{14}\end{pmatrix},\
  \mathbf{f}_3(\pmb{x}) =
    \begin{pmatrix}0\\0\\0\\0\\0\\0\\-2x_8\\2x_7\\-2x_{10}\\2x_9\\-2x_{12}\\2x_{11}\\-2x_{14}\\2x_{13}\\0\\0\end{pmatrix}.
\end{align}
\end{widetext}
If the purity of the qubit is calculated in these coordinates, it turns out to
be
\begin{align}
  P_\text{q} =& \frac{1}{2} + 2\left(x_1 + x_2 - \frac{1}{2}\right)^2\nonumber\\
  & + 2(x_7 + x_{13})^2 + 2(x_8 + x_{14})^2.
\end{align}
This motivates a new choice of coordinates resulting from the
transformation \eqref{eq:z-coord}
\begin{gather}
  \begin{aligned}
  z_1 &= x_1 + x_2 - \frac{1}{2},&  z_5 &= x_7 + x_{13},\\
  z_2 &= x_{12},                 &  z_6 &= x_6 - x_{16},\\
  z_3 &= x_{11},                 &  z_7 &= x_8 + x_{14},\\
  z_4 &= -2 x_1 - x_2 - x_3,     &  z_8 &= x_5 - x_{15},
\end{aligned}
\end{gather}
in which the purity simplifies to
\begin{align}
  P_\text{q} = \frac{1}{2} + 2\left(z_1^2 + z_5^2 + z_7^2\right).
\end{align}
Note that we are left with only eight parameters instead of the original
sixteen. In principle it is possible to consider the complete dynamics by
defining additional parameters $z_9,\dots,z_{16}$, but since the dynamics of
$\pmb{z} = (z_1,\dots,z_8)$ turn out to be closed and we are only interested
in the evolution of the qubit, the other subspace will not be investigated.
\begin{widetext}
The differential equations for the new coordinates read
\begin{align}
  \dot{\pmb{z}} =
  \gamma_1
  \begin{pmatrix}
    0\\-z_2/2\\ -z_3/2\\ -z_1 - z_4 - 3/2\\0\\ -z_6/2\\0\\-z_8/2
  \end{pmatrix}
  + \gamma_2
  \begin{pmatrix}
    0\\-z_2/2\\-z_3/2\\-z_1-z_4-1/2\\0\\-z_6/2\\0\\-z_8/2
  \end{pmatrix}
  + J_1
  \begin{pmatrix}
    2 z_2\\-2 z_1 - z_4 - 1\\0\\0\\z_6\\-z_5\\-z_8\\z_7
  \end{pmatrix}
  + J_2
  \begin{pmatrix}
    2 z_3\\0\\-2 z_ 1- z_4 - 1\\0\\-z_8\\z_7\\-z_6\\z_5
  \end{pmatrix}
  + \alpha
  \begin{pmatrix}
    0\\-2 z_3\\ 2 z_2\\0\\2 z_7\\0\\-2 z_5\\0
  \end{pmatrix}
\end{align}
\end{widetext}

A closer look reveals that the dynamics decouple even further since the
subspaces $(z_1,\dots,z_4)$ and $(z_5,\dots,z_8)$ are independent. The latter
subspace describes the evolution of the coherences of qubit and TLS while the
first one contains the information about the population of the qubit and its
correlations with the TLS.

\noindent\textbf{ACKNOWLEDGMENT}\\
We acknowledge support from the PICS program and from the ANR-DFG
research program COQS (ANR-15-CE30-0023-01, DFG COQS Ko 2301/11-1). C. K acknowledges the support from the 
Volkswagenstiftung Project No. 91004. The work of D. Sugny has been done
with the support of the Technische Universit\"at M\"unchen – Institute for
Advanced Study, funded by the German Excellence Initiative and the European
Union Seventh Framework Programme under grant agreement 291763.

\bibliography{biblio_purification}
\bibliographystyle{apsrev4-1}

\end{document}